\documentclass[final,3p,times,twocolumn]{elsarticle}

\usepackage[T1]{fontenc}
\usepackage[utf8]{inputenc}
\usepackage{amsmath}
\usepackage{amssymb}
\usepackage{arydshln}
\usepackage{epstopdf}
\usepackage{flushend}
\usepackage{hyperref}
\usepackage{graphicx}
\usepackage{lineno}
\usepackage{ulem}
\usepackage{textcomp}
\usepackage{setspace}

\usepackage{miller}

\newcommand{\etal}{\textit{et al.}}
\newcommand{\ie}{\textit{i.e.}}
\newcommand{\eg}{\textit{e.g.}}

\newcommand{\via}{\textit{via}}

\newcommand{\abinitio}{\textit{ab initio}}
\newcommand{\Abinitio}{\textit{Ab initio}}
\newcommand{\insitu}{\textit{in situ}}

\usepackage{xcolor}

\journal{}

\begin{document}
\begin{frontmatter}

\title{Segregation at prior austenite grain boundaries: the competition between boron and hydrogen}

\author[MPIE]{Guillaume Hachet\corref{CA}}
\ead{g.hachet@mpie.de}
\author[MPIE]{Ali Tehranchi}
\author[MPIE]{Hao Shi}
\author[MPIE]{Manoj Prabhakar}
\author[MPIE]{Shaolou Wei}
\author[MPIE]{Katja Angenendt}
\author[MPIE]{Stefan Zaefferer}
\author[MPIE,ICL]{Baptiste Gault}
\author[MPIE,SMPE]{Binhan Sun}
\author[MPIE]{Dirk Ponge\corref{CA}}
\ead{d.ponge@mpie.de}
\author[MPIE]{Dierk Raabe}
\cortext[CA]{Corresponding authors}

\address[MPIE]{Department of Microstructure Physics and Alloys Design, Max-Planck-Institut für Nachhaltige Materialien, 40237, Düsseldorf, Germany}
\address[SMPE]{Key Laboratory of Pressure Systems and Safety, Ministry of Education, School of Mechanical and Power Engineering, East China University of Science and Technology, Shanghai, China}
\address[ICL]{Department of Materials, Royal School of Mines, Imperial College London, Prince Consort Road, London SW7 2BP, UK}

\begin{abstract}
The interaction between boron and hydrogen at grain boundaries has been investigated experimentally and numerically in boron-doped and boron-free martensitic steels using thermal desorption spectrometry (TDS) and \abinitio{} calculations.
The calculations show that boron and hydrogen are attracted to grain boundaries but boron can repel hydrogen. 
This behavior has also been observed using TDS measurements, with the disappearance of one peak when boron is incorporated into the microstructure.
Additionally, the microstructure of both steels has been studied through electron backscattered diffraction, synchrotron X-ray measurements, and correlative transmission Kikuchi diffraction-atom probe tomography measurements. 
While they have a similar grain size, grain boundary distribution, and dislocation densities, pronounced boron segregation into PAGBs is observed for boron-doped steels.
It indicates that boron in PAGBs is responsible for the disappearance of the TDS peaks for the boron-doped steel.
Then, the equilibrium hydrogen concentration in different trapping sites has been evaluated using the Langmuir-McLean approximation. 
This thermodynamic model shows that the distribution of hydrogen is identical for all traps when the total hydrogen concentration is low for boron-free steel. However, when it increases, traps of the lowest segregation energies (mostly PAGBs) are firstly saturated, which promotes failure initiation at this defect type. 
This finding partially explains why PAGBs are the weakest microstructure feature when martensitic steels are exposed to hydrogen-containing environments.
\end{abstract}


\end{frontmatter}

\section{Introduction}
\label{S1}

Hydrogen is a promising carrier molecule for a low-carbon economy and many corresponding industrial projects are under development to use it as an energy carrier or to develop sustainable processes like the direct reduction of iron ore \cite{Souza2021,Ma2022,Odenweller2022,Miocic2023,Cupertino2023}.
It is also known that hydrogen easily diffuses in metals and alloys inducing premature failure of engineering structures, a phenomenon called as hydrogen embrittlement (HE). 
HE is well-known and has been described by numerous models in the literature and thoroughly reviewed in refs. \cite{Robertson2015,Lynch2019,Meda2023,Chen2024}.
However, these models provide only a phenomenological description of HE (\eg{}: hydrogen enhancing the formation of vacancies \cite{McLellan1997,Harada2005,Fukai2006}, localizing plasticity \cite{Beachem1972,Birnbaum1994,Delafosse2001}, reducing the cohesive interface energy of the system \cite{Hondros1996,Oudriss2012} and so on) because the direct observation of hydrogen at the atomic scale is quite challenging given the difficulty to locate hydrogen in alloys experimentally \cite{Takahashi2010,Takahashi2018, Chen2017}.

Steel is one of the candidate materials envisaged to contain and transport hydrogen due to its high mechanical strength, low price, flexibility to form different microstructures from its chemical composition, and the variety of processes that can be applied to the alloys \cite{Bhadeshia2017}.
However, it is also subject to HE \cite{Hirth1980}. 
Different processes have been developed to reduce HE of steels \cite{Barrera2018,Sun2023}, including reducing the grain size \cite{Zan2015}, which is efficient up to a limit, and also modifies other properties of the alloy, potentially to a point of being unsuitable for the envisaged application. 
Steel composition features have also been modified, for instance, to form precipitates and reduce the negative impact of hydrogen by trapping the solute \cite{Depover2015,Chen2020}.
However, once traps inherent to the precipitates, including interfaces with the matrix or sites inside the precipitates, are saturated with hydrogen, they become inefficient for further HE mitigation specifically in typical engineering loading scenarios, which are often characterized by permanent hydrogen exposure.
More recently, the implementation of chemically heterogeneous Mn profiles in the austenite phase of medium Mn-steels has also been proposed, improving the ductility of the alloys in hydrogen \cite{Sun2021}.
However, such treatment is only possible for a relatively narrow range of Mn-steel compositions.

In martensitic steel, hydrogen reduces the cohesion of prior austenite grain boundaries (PAGBs), which is the weakest microstructure feature in steel in a hydrogen environment \cite{Wang2007,Okada2023}. 
A recently proposed solution to mitigate HE in martensitic steels is to favor carbon segregation \cite{Okada2023}. 
In the present work, boron, which is another interstitial solute segregating in PAGBs, should improve the cohesion of the crystalline defect and could mitigate hydrogen embrittlement.
It is an alloying element often incorporated in steel because it improves the hardenability and delays the formation of ferrite \cite{Sharma2019}.
However, the segregation mechanism of boron remained elusive until recently, when it was described using equilibrium segregation and considering the precipitation of borides or boro-carbides at GBs \cite{Prithiv2023,DaRosa2023}.
It has also been observed recently that boron can improve the resistance to HE of steels \cite{Wang2022}, which is also observed in other alloy systems \cite{Chen2022,Bae2023}. 
One explanation theorized from \abinitio{} calculations on $\Sigma 5 (310)$ GB is that boron improves grain boundary cohesion by a strong hybridization between the Fe $s,d$-states, and the B $s,p$-states. 
In contrast, the hydrogen atom exhibits an ion-like character in metals \cite{Wu1994,Kulkov2018} but experimental evidence of this behavior is still missing in the literature.

While the localization and the quantification of hydrogen in metals and alloys remain experimentally challenging, different methods have been developed to detect this element in materials \cite{Koyama2017}.
One of them is thermal desorption spectroscopy (TDS), which allows to quantify the amount of hydrogen located in the microstructure and estimates the trapping energy between hydrogen and different types of crystalline defects \cite{Kissinger1957,Choo1982,Frappart2010,Galindo2017,Drexler2021,Vandewalle2022,Cupertino2023}.
These defects acting as traps for hydrogen are usually identified from the TDS spectra either with complementary experiments \cite{Cupertino2023}, simulations \cite{Counts2010}, or by optimizing the microstructure to favor the formation of certain types of traps \cite{Choo1982,Laureys2020}.
However, discrepancies in binding/interaction energies between hydrogen and crystalline defects have been reported in the literature for martensitic steels \cite{Frappart2010} or even $\alpha$-iron \cite{Bhadeshia2016} depending on the method used.
For instance, the interaction energy of hydrogen in a single vacancy has been estimated from diffusion analysis around -0.55\,eV \cite{Kim1985} while \abinitio{} calculations estimated this energy between -0.30\,eV and -0.25\,eV \cite{Paxton2014}).
Such energy difference makes it difficult to conclude if this crystalline defect in $\alpha$-iron acts as a reversible or an irreversible trap for hydrogen, which is also observed for grain boundaries in martensitic steels \cite{Frappart2010}.
Additionally, martensitic steels contain a complex microstructure with different types of grain boundaries from austenitisation and martensite transformation.
When TDS measurements are performed on hydrogen-charged ultra-low-carbon and high-carbon steels, several temperature ranges have been identified for the desorption of hydrogen from grain boundaries (average and high-angle grain boundaries \cite{Pressouyre1979,Laureys2020,Pinson2021}), making difficult to interpret the hydrogen desorption rate of martensite steels. 

The present work aims to clarify these points by focusing the study on the segregation of hydrogen in boron-doped and boron-free steels.
After a careful investigation of the microstructure of both steels, they have been analyzed through TDS measurements.
Then, \abinitio{} calculations have been performed to understand the interaction between boron and hydrogen at grain boundaries.
Finally, the equilibrium concentration of hydrogen has been evaluated from TDS measurements using the Langmuir-McLean approximation to determine the hydrogen partitioning in the microstructure when the two steels are hydrogen-charged.

\section{Microstructure characterization}
\label{S2}

\subsection{Experimental details}
\label{S21}

We used two alloys named LC and B-LC steels in this work and their chemical compositions are Fe-C$_{0.15}$-B$_{0.0005}$-Mn$_{1.50}$-Si$_{0.37}$ (wt\%) and Fe-C$_{0.146}$-B$_{0.0025}$-Mn$_{1.46}$-Si$_{0.40}$ (\%wt.), respectively \cite{SukumarPhD}.
All steels have been homogenized at an austenitization temperature of 1373\,K for 30 seconds and then quenched using helium gas in a Bähr DIL805 dilatometer with a cooling rate of 200\,K.s$^{-1}$. 
The microstructures of LC and B-LC have been characterized through electron backscatter diffraction (EBSD), synchrotron X-ray experiments, and correlative transmission Kikuchi diffraction (TKD)-atom probe tomography (APT) to quantify all differences between the microstructure of the two steels.

The EBSD maps have been acquired using a high-resolution field emission GEMINI SEM 450 (Carl Zeiss Microscopy) equipped with a Hikari XP (Ametek/EDAX, USA) EBSD detector. 
An acceleration voltage of 15\,kV, probe current of 2\,nA, and step size of 0.1\,$\mu$m have been chosen for all mappings.
The post-processing of all images has been carried out using the TSL OIM Analysis v8 on pixels with a confidence index above 0.1 on a surface of at least 1.5 $\times 10^4$ $\mu$m$^{2}$.
Then, the MTEX 5.11.1 software toolbox was used to identify the misorientation angle of all grain boundaries and to reconstruct PAGBs from the EBSD maps with MATLAB R2021b \cite{Bachmann2010}.

The microstructure has been analyzed using synchrotron X-ray diffraction measurements.
They were conducted at Deutsches Elektronen-Synchrotron (DESY, Hamburg, Germany) on the Petra III P-02.1 beamline at 60\,keV.
A high-energy transmission X-ray beam with a wavelength of 0.207381 \AA{} was shed on square-shaped specimens (10$\times$10$\times$1\,mm$^{3}$) to collect two-dimensional diffractograms at a working distance of 969\,mm.  
Before conducting quantitative diffraction analyses, the instrumental parameters have been calibrated using the diffraction patterns of NIST standard LaB$_6$.
All recorded two-dimensional diffractograms have been post-processed using the GSAS-II software \cite{Toby2013}.

Finally, correlative TKD-APT measurements have been performed on PAGBs of LC and B-LC to quantify the segregation of boron.
The sample has been prepared using a dual-beam SEM-focused ion beam (FIB) instrument (FEI Helios Nanolab 600i) using an \insitu{} lift-out procedure.
Prior to conducting the APT measurements, specimens were analyzed using TKD to characterize the grain boundaries with a Digiview V) EBSD detector on MERLIN SEM (Zeiss Microscopy).
The prepared APT specimens have been investigated in a CAMECA LEAP 5076XS instrument, operated in laser mode at 60\,K with a pulse rate of 200\,kHz, pulse energy of 30\,pJ, and a detection rate of 50 ions per 1000 pulses.
The three-dimensional reconstructions have been performed using the AP suite 6.3 software.

\subsection{Microstructure evolution with boron addition}
\label{S22}

\begin{figure*}[ht]
\centering
\includegraphics[width=0.8\linewidth]{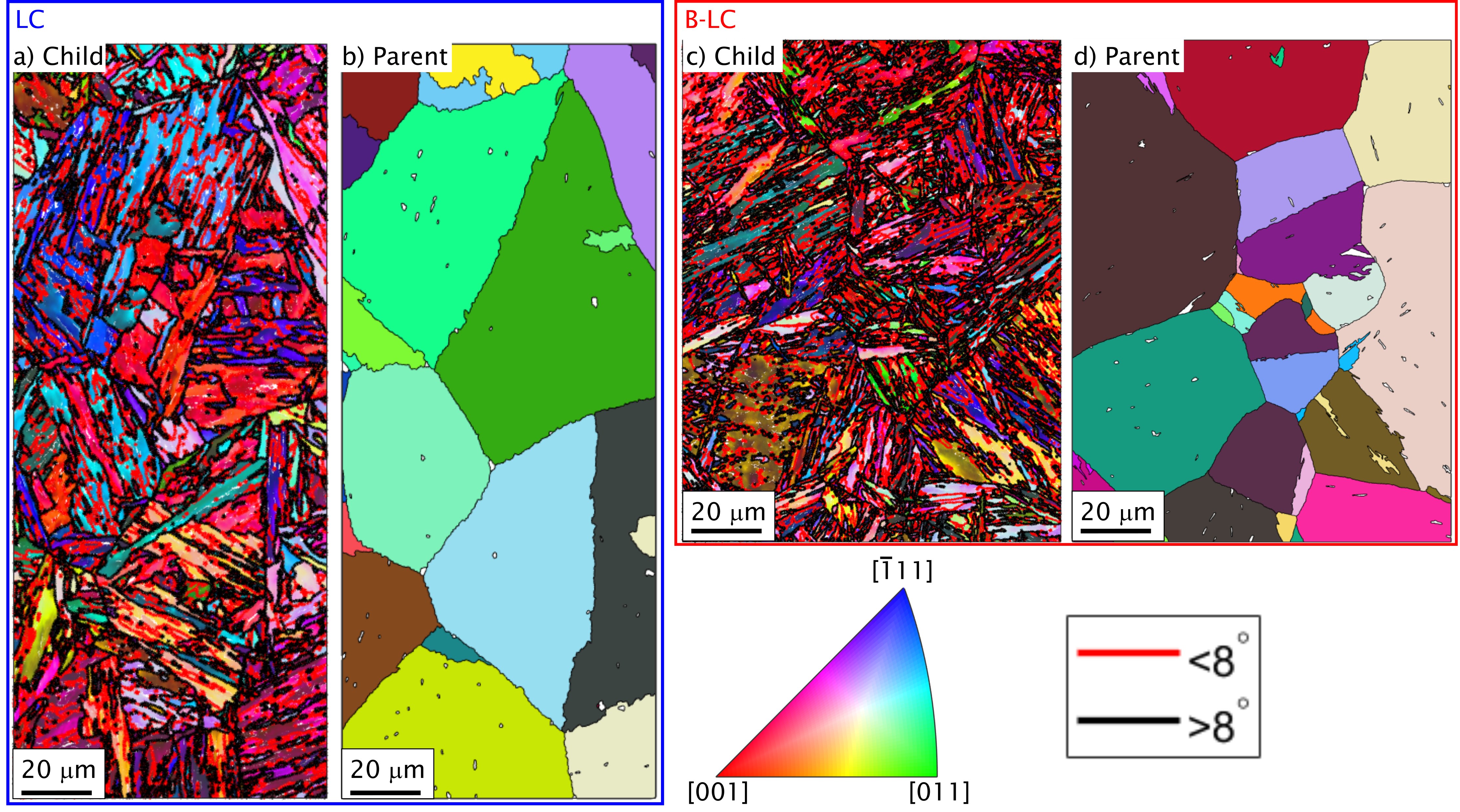}
\caption{In plane IPF maps of a) LC and c) B-LC, and d) its corresponding reconstructed parent grains.
The SGBs having a dislocation-like microstructure (with a misorientation lower than 8°) are represented in red and the MMBs with the KS variant misorientation and PAGB are represented in black in figs. a and c. 
The PAGBs are also represented in black in figs. b and d.}
\label{FigLengthGB}
\end{figure*}

The microstructure evolution with boron addition is first investigated in this work by comparing the difference of grain size and grain boundary distribution in boron-doped (B-LC) and boron-free (LC) steels. 
From the inverse pole figure (IPF) maps provided in figs \ref{FigLengthGB}.a and \ref{FigLengthGB}.c of LC and B-LC steels, both have a martensite microstructure after quenching.
The grain size has then been estimated from the IPF maps presented in \ref{AppMicrostructure}, is 15.4\,$\mu$m $\pm$ 10.8\,$\mu$m for LC and 11.3\,$\mu$m $\pm$ 10.4\,$\mu$m for B-LC.
These diameters have been determined by calculating the area fraction of each grain assuming they are a circle.
A slight difference in the grain size is observed between LC and B-LC but remains neglectable given the large fluctuations for both steels.
The reconstructed parent grains of figs \ref{FigLengthGB}.a and \ref{FigLengthGB}.c
are presented in figs \ref{FigLengthGB}.b and \ref{FigLengthGB}.d, respectively.
They have been obtained by regrouping grain boundaries following the Kurdjumov-Sachs (KS) orientation relationship (OR), \ie{}: regrouping the $(111)_\gamma \parallel (\overline{1}01)_\alpha$  and $\lbrack 011 \rbrack_\gamma \parallel \lbrack \overline{1}\overline{1}1 \rbrack_\alpha$ \cite{Morito2003a,Morito2005,Morsdorf2015,Bhadeshia2017,Engler2024}, following the instruction provided by MTEX \cite{Niessen2022}.

Further, three categories of grain boundaries have been sorted for LC and B-LC: (i) prior austenite grain boundaries that do not follow KS OR, (ii) martensite-martensite boundaries (MMBs) with a relevant KS OR, and (iii) subgrain boundaries (SGBs) having a misorientation lower than 8°.
Table \ref{TabGBsize} presents the length distribution in the fixe  of these three types of boundaries ($l_{PAGB}$, $l_{MMB}$, and $l_{SGB}$) for both steels. 
Both microstructures contain mostly grain boundaries from martensite transformation (more than 90\% of GBs are SGBs and MMBs for LC and B-LC).
The distribution of grain boundary types in LC and B-LC indicates a similar microstructure with boron incorporation and suggests that boron has a minor impact on the grain boundary distribution.

\begin{table}[ht]
\caption{\label{TabGBsize} Summary of the length distribution of PAGBs ($l_{PAGB}$), MMBs following KS OR ($l_{MMB}$), and SGBs ($l_{LSGB}$) in LC and B-LC. 
They are the sum of segments' length from the map shown in fig. \ref{FigLengthGB}.}
\centering
\begin{tabular}{l|ccc}
\hline
 \ & $l_{PAGB}$ mm (\%) & $l_{MMB}$ mm (\%) & $l_{SGB}$ mm (\%) \\
\hline
LC & 1.92 (8.3) & 10.09 (43.5) & 11.69 (48.2) \\
B-LC & 2.15 (7.5) & 15.68 (47.9) & 14.94 (45.6) \\
\end{tabular}
\end{table}

\begin{figure}[ht!]
\centering
\includegraphics[width=0.9\linewidth]{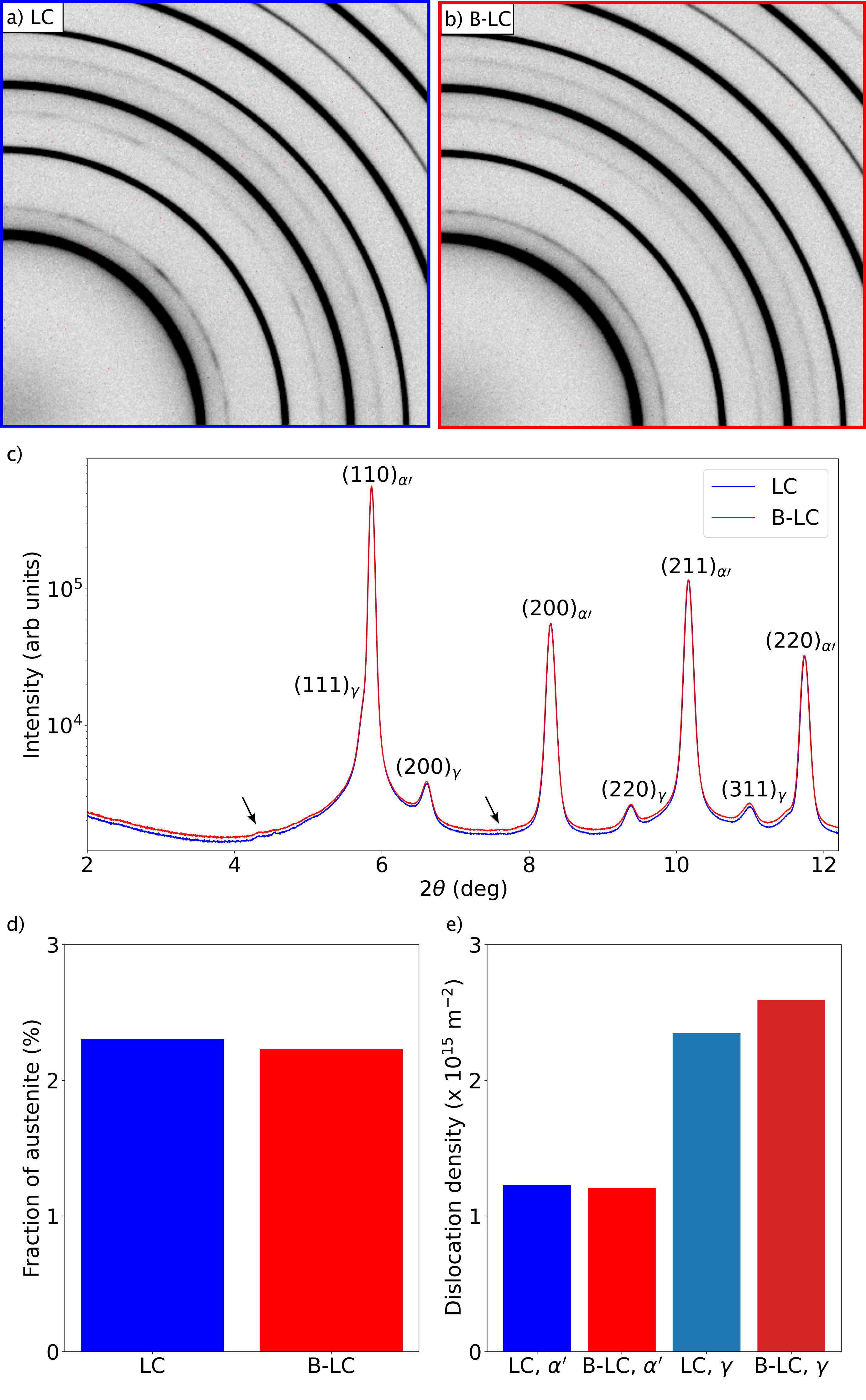}
\caption{Snapshots of two-dimensional diffractograms acquired for a) LC and b) B-LC steels.
c) Circular integration of the two-dimensional diffractograms (in logarithmic scale).
The major diffraction peaks are attributed to martensite ($\alpha^{\prime}$) and retained austenite ($\gamma$) phases.
The arrows highlight fluctuations of the signal attributed to carbides.
d) Austenite phase fraction ($f_{\gamma}$) obtained for LC and B-LC 
e) Dislocation density in $\alpha^{\prime}$ and $\gamma$ for both LC and B-LC steels.}
\label{FigSynchrotronXray}
\end{figure}

The synchrotron X-ray measurements show similar two-dimensional diffractograms for LC and B-LC (figs. \ref{FigSynchrotronXray}.a and \ref{FigSynchrotronXray}.b, respectively). 
Peaks from the martensite microstructure and the retained austenite are obtained during the measurements when the circular integration of the two-dimensional diffractogram is presented in logarithmic scale.
This austenite phase should be located at lath martensite boundaries according to the literature for medium carbon steels \cite{Sherman2007,Morito2011} and was not detected during the EBSD analyses.
It is either because it is unstable and transforms to martensite when it reaches the free surfaces during metallographic sample preparation or the thickness of the austenite layer is too thin to be detected by EBSD. 
When the synchrotron X-ray diffraction profiles are integrated, the contribution related to the austenite phase becomes clearer (fig. \ref{FigSynchrotronXray}.c).

Fig \ref{FigSynchrotronXray}.d presents the volume fraction of the $\gamma$ phase ($f_{\gamma}$), which is 2.26$\pm$0.04\,\% for both steels and is deduced by integrating peaks related to the martensite and austenite phases.
This method was chosen over the Rietveld refinement analysis approach because the latter neglects fluctuations that are observed at 4.5° and 7.5° (highlighted with black arrows in fig. \ref{FigSynchrotronXray}.c). 
These fluctuations are related to carbides, which could be either cementite, $\eta$-carbides, or $\varepsilon$-carbides according to the literature \cite{Allain2018} that some are observed through ECCI experiments (fig \ref{FigMicrostructure}.d).

Then, the crystalline size $D$ and dislocation density $\rho$ of both phases have been determined using the Williamson-Hall approach \cite{Williamson1954,Wei2023}:
\begin{equation}
    \beta \cos\left(\theta_{hkl}\right)= \frac{\lambda}{D} + \varepsilon_{micro}\sin\left(\theta_{hkl}\right),
\label{eqFWHM}
\end{equation}

with $\beta$ the full width at half maximum of the diffraction peak at $\theta_{hkl}$, $\lambda$ the wavelength of the beam ($\lambda$ = 0.207381 \AA), $\varepsilon_{micro}$ the micro-strain, and $\theta_{hkl}$ the position of the $\lbrace hkl \rbrace$ reflection group.
The dislocation density $\rho$ can then be estimated using \cite{Williamson1954,Wei2023}:
\begin{equation}
    \rho = \frac{2\sqrt{3}\varepsilon_{micro}}{bD},
\label{eqrho}
\end{equation}

with $b$ being the magnitude of the Burgers vector of the screw dislocation in the martensite phase (2.48 \AA) and of the edge dislocation in the austenite phase (2.54\,\AA).
Following this approach, a similar dislocation density has been obtained in LC and B-LC (fig. \ref{FigSynchrotronXray}.d) of 1.21 $\pm$ 0.01 $\times$ 10$^{15}$ m$^{-2}$ in the martensite phase and 2.4 $\pm$ 0.1 $\times$ 10$^{15}$\,m$^{-2}$ in the austenite phase, respectively.
From these measurements, boron incorporated in the steels has a minor effect on the fraction of retained austenite and dislocation density in both phases.

\begin{figure*}[ht!]
\centering
\includegraphics[width=0.9\linewidth]{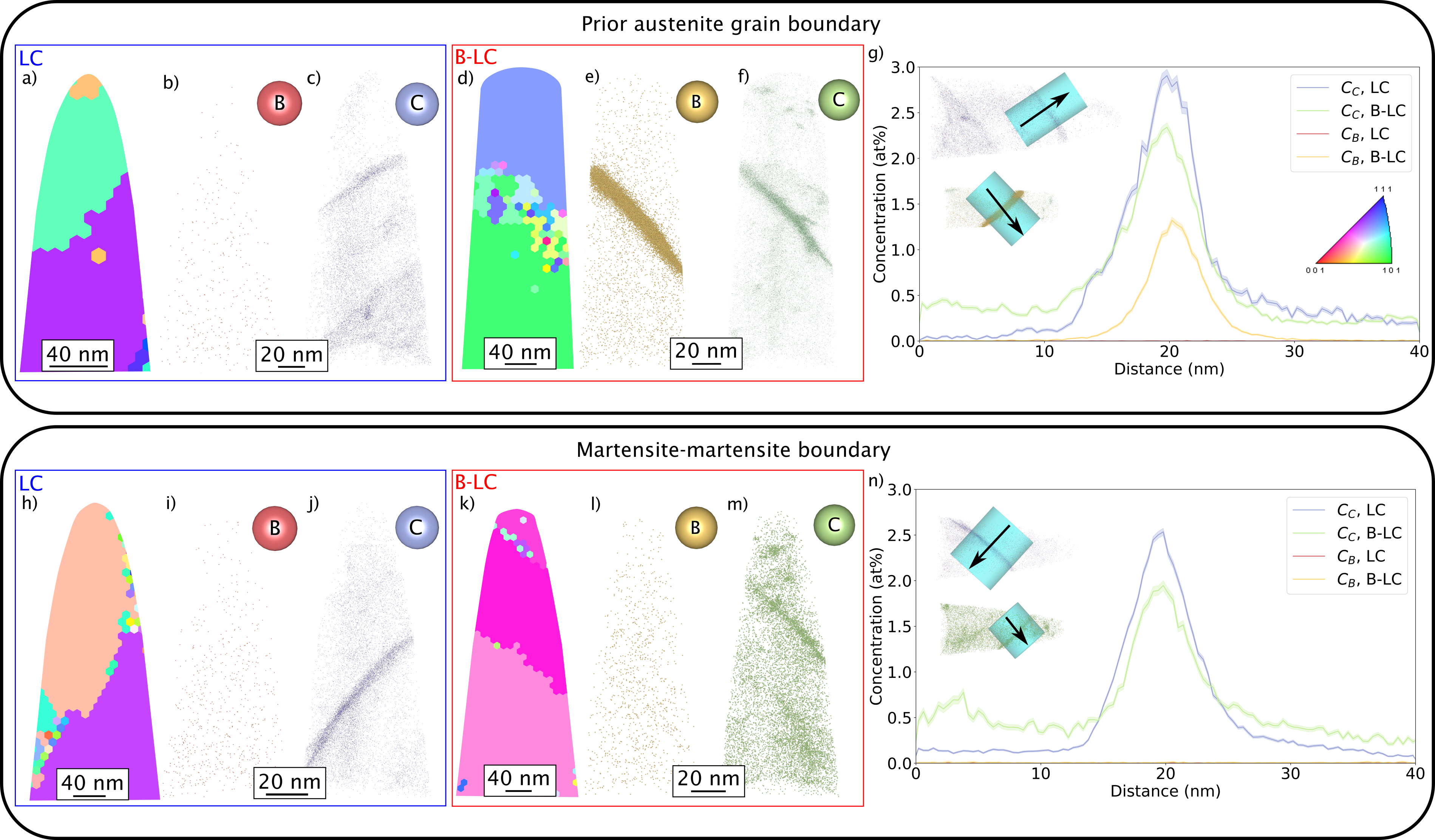}
\caption{Correlative TKD-APT analysis of PAGBs (a-g) and martensite-martensite boundaries (h-n) in LC and B-LC steels. 
a) (and h) IPF map of LC with the three-dimensional reconstruction of the APT needle showing b) (and i) boron (red atoms) and c) (and j) carbon (blue atoms).
d) (and k) IPF map of B-LC with the three-dimensional reconstruction showing e) (and l) boron (orange atoms) and f) (and m) carbon (green atoms).
g) (and n) Corresponding one-dimensional composition profile in PAGBs (in MMBs).}
\label{FigAPT}
\end{figure*}

Finally, correlative TKD-APT experiments have been performed to characterize the boron segregation at PAGBs and MMBs.
A site-specific lift-out has been performed for localizing the different grain boundaries from LC and B-LC microstructures, following the procedure presented in previous work \cite{Prithiv2023} before the TKD observations.
Figs \ref{FigAPT}.a and \ref{FigAPT}.d provide the TKD maps of PAGBs and figs \ref{FigAPT}.h and \ref{FigAPT}.k of MMBs from LC and B-LC before the APT measurements, respectively. 
The orientation relationship of the different GBs has first been checked to confirm the nature of each boundary. 
For PAGB, the misorientation angle/axis between the grains is 44°/[0.526 -0.316 -0.789] for the TKD map of fig \ref{FigAPT}.a and is 59°/[0.539 -0.539 0.647] for the TKD map of fig \ref{FigAPT}.h.
The latter orientation is close to the variant V1 obtained from the KS OR (60°/[0.577 -0.577 0.577]) \cite{Morito2003b,Engler2024}, indicating that it is a MMB while the former is a PAGBs.
For B-LC, it is 44°/[-0.743 0.656 -0.131] for the TKD map of fig \ref{FigAPT}.d and it is 52°/[-0.734 -0.587 0.342] for the TKD map of fig \ref{FigAPT}.k.
The latter grain boundary of B-LC is close to the variant V17 obtained from the KS OR (51.7°/[-0.659 0.363 -0.59]) \cite{Morito2003b,Engler2024}, it suggests that it is a MMB and while the grain boundary of fig \ref{FigAPT}.d can be considered as a PAGB.

Figs \ref{FigAPT}.b, c present three-dimensional reconstructions showing only boron and carbon atoms for LC while figs. \ref{FigAPT}.e, f present three-dimensional reconstructions of the same elements in B-LC.
Single events have been separated from multiple events to improve the visualization of boron and carbon in the reconstructed data.
This data filtering has been shown to provide an improved signal-to-background ratio for species in low concentration in Fe-based materials \cite{Yao2010}.
More details are given at \ref{AppAPT}, showing an improved boron quantification in an interface obtained from martensitic transformations.
Figs \ref{FigAPT}.i, j and figs. \ref{FigAPT}.l, m present a similar reconstruction but for a MMB for LC and B-LC, respectively.
For all reconstructions, the grain boundary can be localized from the pronounced carbon segregation.
Figs \ref{FigAPT}.g and \ref{FigAPT}.n plot the resulting composition profile of the interface marked by a blue cylinder in the corresponding APT reconstruction shown inset.
While the concentration of carbon is similar for all interfaces (maximum concentration around 2.5 at.\%), the maximum boron concentration is less than 0.1 at.\% for LC and higher than 1.2 at.\% for B-LC in PAGBs.
However, this pronounced segregation is not observed in MMBs of B-LC.   
This difference highlighted the pronounced segregation of boron in PAGBs for B-LC, which is not observed for LC, or in MMBs of B-LC, which contains almost no boron.

In this section, we described the microstructure obtained in both LC and B-LC after quenching from 1100°C in He gas (cooling rate of $\sim$200\,K.s$^{-1}$). 
A similar microstructure is observed for LC and B-LC with an average grain size around 13 $\mu$m (when using the grain size of spherical grains) for both steels with the appearance of carbides, confirmed through synchrotron X-ray experiments.
Additionally, these experiments show that 2\,vol.\% of retained austenite is observed in these steels.
This fraction of retained austenite is lower than the fraction deduced in medium carbon steel reported in the literature (which is 5.5\%) \cite{Sherman2007}.
The dislocation density in the martensite phase has been estimated from the synchrotron X-ray experiments as 1$\times$10$^{15}$\,m$^{-2}$, and agrees with measurements performed in the literature \cite{Morito2003b}.
In the austenite phase, a higher dislocation density has been estimated and is 3$\times$ 10$^{15}$\,m$^{-2}$.
Finally, the APT measurements correlated with TKD analysis reveal a major change in behavior between boron-doped and boron-free steels, which is the segregation of boron to the PAGBs. 
At these interfaces, the maximal boron concentration is $\sim$1\,at.\% but can increase up to 8\,at.\% according to previous work \cite{Prithiv2023}.
In addition, previous secondary ion mass spectroscopy analyses had demonstrated that the segregation at other martensite boundaries is minor \cite{Prithiv2023}. 
Since it is expected to observe the same solute concentration for LC and B-LC at these interfaces, the trapping behavior towards hydrogen should be similar at these interfaces.
Further, hydrogen is inserted into these microstructures and detected using TDS measurements.
According to the microstructure characterization, hydrogen could diffuse and be potentially trapped in vacancies, dislocations from martensite transformation, grain boundaries (martensitic laths, block and packet boundaries, PAGBs), precipitates (carbides), and in the austenite phase.

\section{Impact of boron segregation on the hydrogen trapping behavior}
\label{S3}

\subsection{Experimental details}
\label{S31}

The incorporation of hydrogen into $LC$ and $B-LC$ has been performed electrochemically, using a three-electrode system on 10$\times$10$\times$1\,mm$^{3}$ samples.
The procedure consisted of imposing a current of -25\,mA.cm$^{-2}$ for 1\,h in a 0.1\,M $H_2SO_4$ aqueous solution with 0.3\,wt\% of NH$_4$SCN at room temperature.  
After charging, the oxide layers were removed from the sample using 600 SiC grinding paper before the TDS measurements.
For all conditions, the time between the end of the electrochemical charging and the beginning of the temperature ramping is lower than 5\,min, which is performed in an infrared furnace from 298\,K to 1173\,K with a mass spectrometer using the G8 GALILEO apparatus from Bruker$^{\circledR}$.
The heating rate $\varphi$ varied between 8\,$K.min^{-1}$ and 32\,$K.min^{-1}$ to locate where hydrogen is trapped in the microstructure.

\subsection{Results}
\label{S32}

\begin{figure*}[ht!]
\centering
\includegraphics[width=0.99\linewidth]{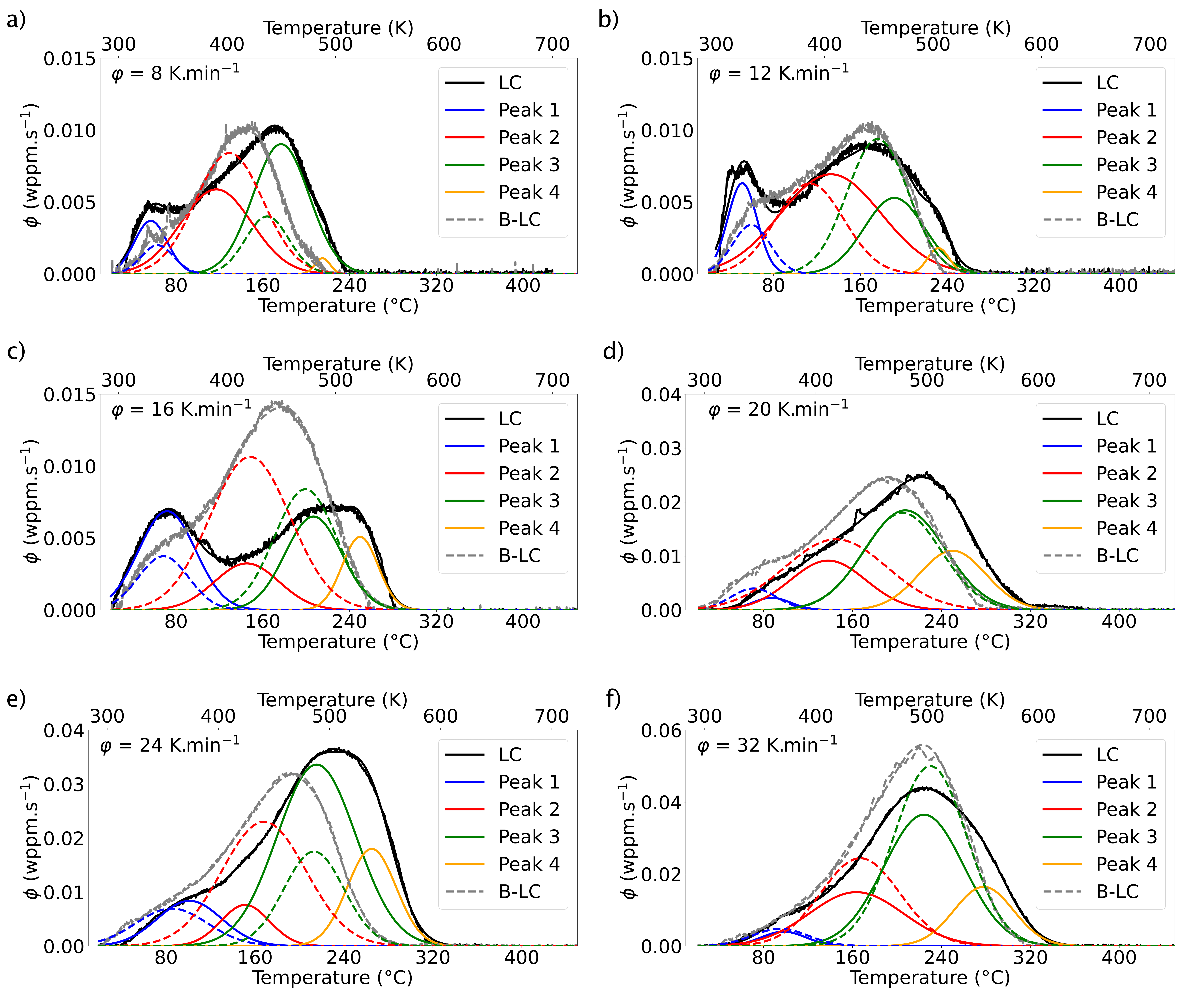}
\caption{TDS spectra of LC and B-LC heated for different temperature ramping at  a) 8\,K.min$^{-1}$, b) 12\,K.min$^{-1}$, c) 16\,K.min$^{-1}$, d) 20\,K.min$^{-1}$, e) 24\,K.min$^{-1}$, and f) 32\,K.min$^{-1}$.
The sum of all deconvoluted curves is represented for all spectra to compare the difference between the fit and the experiment.}
\label{FigTDSSpectra}
\end{figure*}

The TDS spectrum of each condition is given in fig. \ref{FigTDSSpectra} with the hydrogen desorption rate $\phi$ not represented between 723\,K and 1173\,K because the signal is flat.

The absolute difference between the TDS spectra and a function being the sum of several Gaussian distributions is then minimized using a conjugate-gradient algorithm to determine the number of peaks for each spectrum.
Following this procedure, four peaks are detected in LC steel and only three in B-LC for all $\varphi$.
Boron mostly segregating into PAGBs \cite{Sharma2019,Prithiv2023,DaRosa2023}, the last peak should correspond to hydrogen trapped in this crystalline defect in $LC$ steels.
This result agrees with previous studies that have shown that the desorption of hydrogen located in HAGBs at temperature ranges higher than the desorption of hydrogen located in dislocations \cite{Pinson2021,Pressouyre1979}.
In these studies, the trapping energy of hydrogen in HAGBs was estimated between -0.61\,eV and -0.55\,eV, whereas the trapping energy of hydrogen in average grain boundaries was -0.27\,eV \cite{Pressouyre1979}.
However, according to the hydrogen desorption behavior observed in LC and B-LC of this work, the difference between these two trapping energies may be more related to PAGBs and martensite boundaries (MMBs and SGBs) than their misorientation angle. 
PAGBs are GBs existing at high temperatures (1100\,°C) whereas martensite boundaries are formed in a parent austenite grain at lower temperatures (the start of the martensite transformation is below 450\,°C), following the KS OR.
Consequently, the coherency of PAGBs is expected to be different from martensite boundaries which can modify the trapping behavior of hydrogen even though both types of boundaries have low and high misorientation angles.

Further, the identification of the possible trapping sites of the different peaks is made following previous work \cite{Choo1982,Laureys2020,Pinson2021}.
In these works, the microstructure of $\alpha$-iron and ultra-low carbon steel has been optimized to favor the formation of certain types of traps.
The segregation energy of hydrogen in different traps, deduced from the peak shift is also compared with results from the literature \cite{Frappart2010,Bhadeshia2016,Galindo2017} to determine the possible trapping sites associated with each peak.

The variation of $\varphi$ produces a shift of the different peaks.
Using Kissinger’s theory, the trapped energy of hydrogen with a defect $E^{\rm tra}_{\rm H-def}$ can be determined using \cite{Kissinger1957,Choo1982,Frappart2011,Kirchheim2016,Matson2021}:
\begin{equation}
	\frac{E^{\rm seg}_{\rm H-def}}{k_{\rm B}} = \frac{E^{\rm D}_{\rm H} + E^{\rm tra}_{\rm H-def}}{k_{\rm B}} = \frac{\partial\ln(\varphi/T_{pj})}{\partial(1/T_{pj})},
\label{eqKissPlot}
\end{equation}
where $k_{\rm B}$ being the Boltzmann constant, $E^{\rm seg}_{\rm H-def}$ the segregation energy of hydrogen in a defect, and $E^{\rm D}_{\rm H}$ the energy barrier for hydrogen to diffuse in lattice ($E^{\rm D}_{\rm H}$= -0.043\,eV \cite{Kirchheim2016,Borges2022}).

\begin{figure}[ht!]
\centering
\includegraphics[width=0.9\linewidth]{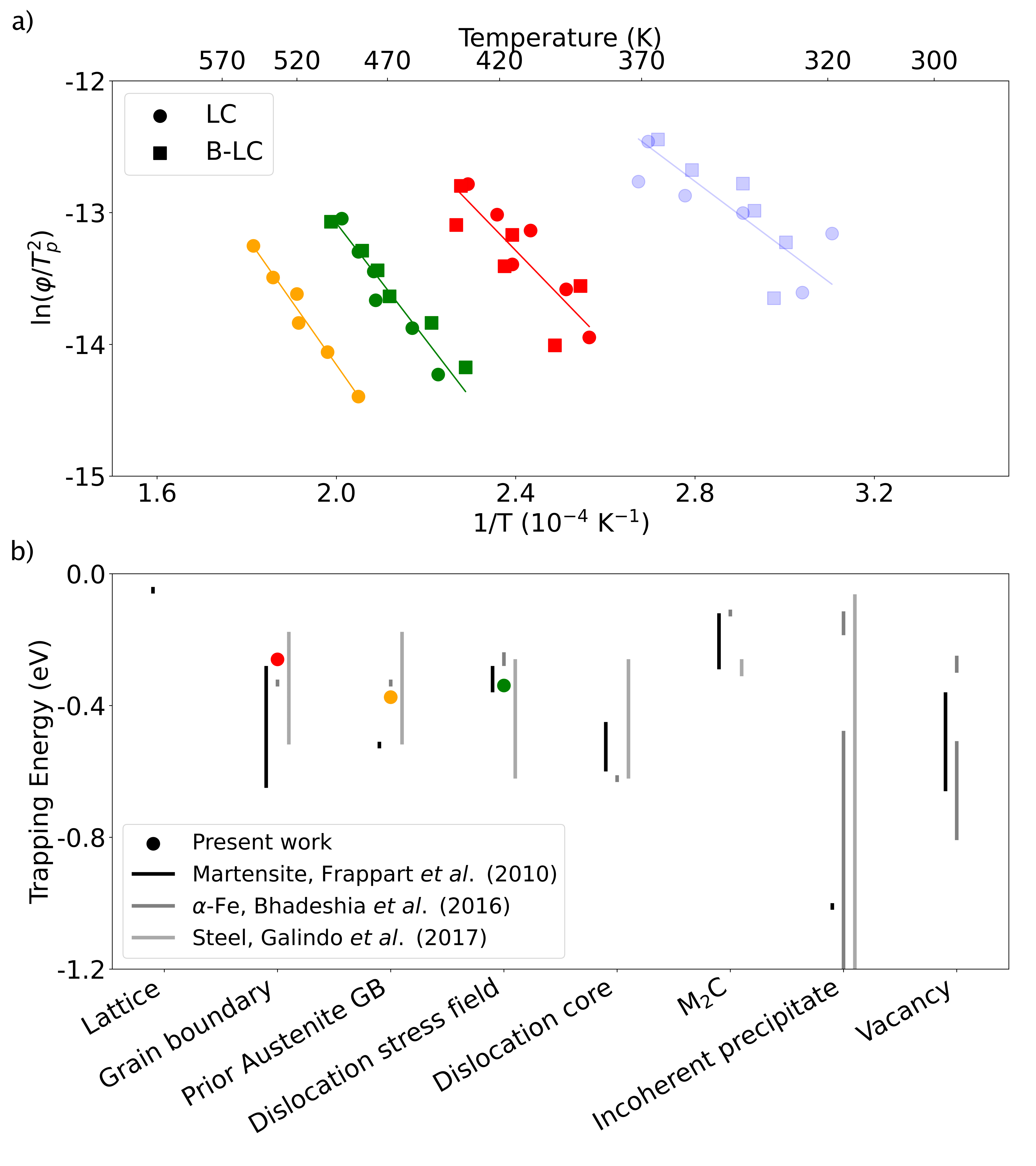}
\caption{Trapping energy deduced from the TDS spectra.
a) Associated Kissinger plot for $LC$ and $B-LC$ steels.
b) Trapping energies between hydrogen and different microstructural features in $\alpha-Fe$ \cite{Bhadeshia2016}, steel \cite{Galindo2017} and martensite steel \cite{Frappart2010} compared to the energies determined in this work.
The shadowed points and curve correspond to the first peak which gives an inaccurate segregation energy.}
\label{FigTDSETrap}
\end{figure}

Fig. \ref{FigTDSETrap}.a presents the Kissinger plot of the different peaks shown in fig. \ref{FigTDSSpectra} for LC and B-LC steels.
The first peak from the TDS spectra of fig. \ref{FigTDSSpectra} should correspond to mobile hydrogen in the lattice (from the slope of the blue curve). 
The energy obtained is -0.16 eV and is lower than the theoretically expected activation barrier for diffusion of hydrogen in martensitic steel or $\alpha$-iron, which is between -0.04\,eV and -0.1\,eV \cite{Borges2022,Frappart2010,Kirchheim2016}.
The main reason is that the determination of this energy comes from the temperature at which the hydrogen desorption rate $\phi$ is the highest.
When the heating rate ($\varphi$) is low, this temperature should come around or below room temperature \cite{Laureys2020}, the starting temperature from our experiments.
The second peak (red curves) has been attributed to hydrogen trapped at GBs in previous work \cite{Choo1982,Laureys2020}. 
Our results show a fourth peak which pertains to hydrogen desorbing from PAGBs.
Consequently, this second peak should mostly be related to the desorption of hydrogen trapped in grain boundaries from martensite transformation, which follows the KS OR, and has the same parent austenite grain.
The trapping energy deduced from our experiment is -0.26\,eV for hydrogen segregating into this type of trapping site (deduced from the slope of the red curve).
This energy is in the range of the ones reported in the work of Galindo \etal{} \cite{Galindo2017}, but outside the range given in Frappart \etal{} \cite{Frappart2010,Frappart2011} (fig. \ref{FigTDSETrap}.b).
The slope of the green curve gives the average trapping energy of -0.34\,eV, and should represent mostly hydrogen in dislocations according to the literature \cite{Choo1982,Laureys2020}. 
This value is in the range of the energies given in previous studies \cite{Frappart2010,Bhadeshia2016,Galindo2017}.
The last peak, observed only for the boron-free steel (orange curve in fig. \ref{FigTDSETrap}.b) should be hydrogen in PAGBs with a trapping energy of -0.37\,eV. 
This value is close to one reported in the work of Bhadeshia \cite{Bhadeshia2016} but higher than the one reported in the work of Frappart \etal{} \cite{Frappart2010}.

In addition, hydrogen could also be trapped in the 2\% of retained austenite observed in LC and B-LC steels and in or at carbides \cite{Bhadeshia2016,Galindo2017,Pinson2021}, precipitates observed by ECCI and synchrotron X-ray measurements.
When hydrogen is trapped in austenite, a peak from the TDS spectra above 573\,K should be obtained \cite{Pinson2021}, which is not the case, neither for LC nor for B-LC.
Different possible hypotheses could explain this behavior and are presented further.
The austenite, which is too compressed to transform into martensite, is becoming less attractive site for hydrogen to segregate, even at the interface with the matrix its concentration becomes too low.
Such effect is observed in the case of edge dislocations for BCC and FCC metals, where the stress field of the compressed area induces a negative binding energy for hydrogen, and favors its segregation into the tensile area of the defect \cite{Tang2012,Bhatia2014}.
Another possible explanation is that the thin layer of this phase reduced the temperature range for hydrogen desorption from it and the small fraction makes it impossible to observe an apparent desorption peak by TDS measurements.

A similar observation is deduced for carbide: its concentration is too low to observe an apparent peak by TDS measurements.
A deconvolution of the TDS spectra considering these defects could be done near peaks 2 and 3, but might lead to an inaccurate overfit.
Another peak has been found for hydrogen in microvoids in previous works, which has an amplitude significantly lower than the other peaks \cite{Choo1982,Laureys2020}.
However, no additional peak has been observed in either of the two steels. 
One possible explanation is that hydrogen recombines in these microvoids, making it impossible for hydrogen to desorb at the temperature range studied in this work \cite{Laureys2020}.

To summarise, an additional peak has been observed when performing TDS measurements for different heating rates $\varphi$ on boron-free steel compared to boron-doped steel.
This peak has been attributed to HAGBs in previous studies \cite{Pressouyre1979,Pinson2021}.
In the previous section, we observed that both LC and B-LC steels have similar microstructure, except for a more pronounced segregation of boron into PAGBs.
Further, \abinitio{} calculations are performed for a better understanding of the interaction between boron and hydrogen in PAGBs.

\section{Interaction energy of hydrogen and boron at grain boundaries}
\label{S4}

\subsection{Computational details}
\label{S41}

\begin{figure}[ht!]
\centering
\includegraphics[width=0.9\linewidth]{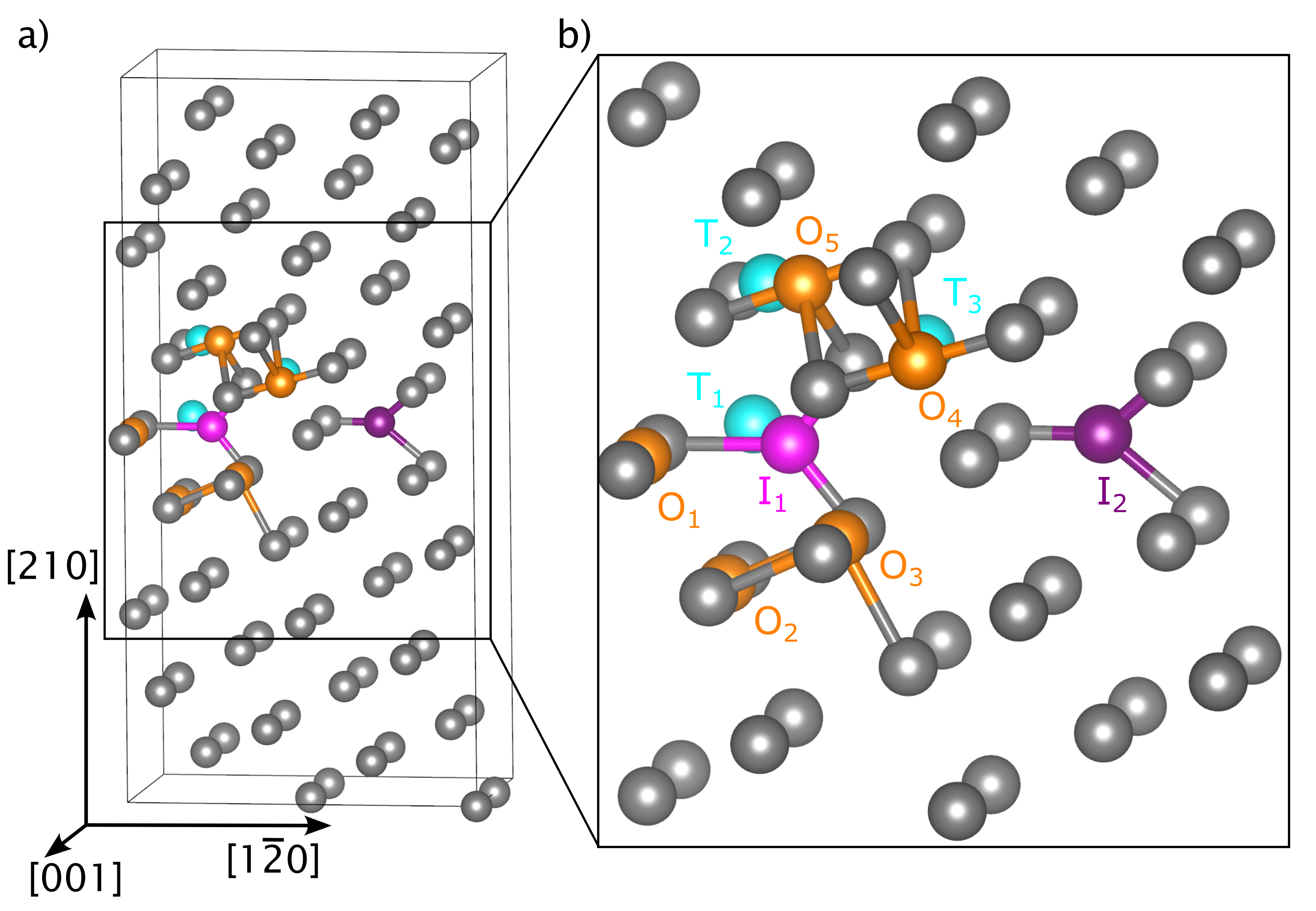}
\caption{Structure of $\alpha$-Fe$\Sigma 5(210)$ GB supercell studied in this work with the different positions where $B$, $C$, and $H$ have been inserted.
a) Full supercell structure. 
b) Zoom in the core of the grain with the octahedral sites ($O_1$, $O_2$, $O_3$, $O_4$ and $O_5$) represented in orange, the tetrahedral sites ($T_1$, $T_2$ and $T_3$) in cyan, and interstitial sites ($I_1$ and $I_2$) represented in shade of magenta.  
The grey atoms represent Fe.}
\label{FigAbinitioCalcSites}
\end{figure}

While hydrogen can segregate into any crystalline defects, including PAGBs, boron is only in high concentration in PAGBs.
Therefore, the difference in behavior on the hydrogen desorption rate described with TDS measurements should be due to the interaction between hydrogen and boron at this crystalline defect.
Consequently, \abinitio{} calculations have been performed to understand this interaction by determining the interaction energies between solutes and the structural defect.
have been carried out using the Vienna \Abinitio{} Simulation Package (VASP) code \cite{Kresse1993, Kresse1996a,Kresse1996b} in the generalized gradient approximation (GGA) using the PBE functional \cite{Perdew1996}.
A plane-wave cut-off of 500\,eV has been taken for all calculations. 
The convergence tolerance of the atomic force is 0.01\,eV/{\AA} and the Brillouin-zone integration was made using Methfessel–Paxton smearing \cite{Methfessel1989}.
The magnetic state of $\alpha$-Fe is ferromagnetic and a $4\times 4\times 4$ supercell containing 128\,Fe atoms has been used for the bulk calculations.
For the reciprocal space integration, a $\Gamma$-centered Monkhorst-Pack~\cite{Monkhorst1976} \textit{k}-point sampling of $8\times 8\times 8$ has been used for bulk $\alpha$-Fe calculations.
The lattice constant and magnetic moments of the ferromagnetic $\alpha$-Fe were computed as 2.834\,{\AA} and 2.2\,$\mu_{\rm B}$/atom, respectively, in agreement with the previous studies \cite{Zhang2015}.  
The supercell containing the $\alpha$-$\Sigma$5(210)[001] grain boundary consisted of 80 Fe-atoms.  
For the reciprocal space integration, a $\Gamma$-centered Monkhorst-Pack~\cite{Monkhorst1976} \textit{k}-point sampling of $5\times 10\times 16$ has been used to ensure convergence.
The solute atoms H, C, and B have firstly been incorporated in different locations of the defect (in interstitial sites $I_1$ and $I_2$), represented in fig. \ref{FigAbinitioCalcSites}. 
Then, additional calculations have been performed to describe the interaction between a boron-doped GB with an additional solute atom.
In this situation, boron has been incorporated in $I_1$ and a second solute atom (hydrogen and boron) has been incorporated in interstitial site $I_2$ or in interstitial sites in the vicinity of the GB (in octahedral and tetrahedral sites).

The interaction energy between a solute and the Fe$\Sigma 5(210)$ GB ($E^{\rm inter}_{\rm X-GB}$) has been determined using: 
\begin{align}
    E^{\rm inter}_{\rm X-GB}=(E^{\rm sp}_{\rm X-GB}-E^{\rm sp}_{\rm GB})-(E^{\rm sp}_{\rm X}-E_{\rm bulk})
\label{eqEinterDFT}
\end{align}
where $E^{\rm sp}_{\rm X-GB}$, $E^{\rm sp}_{\rm GB}$ are the energies of the supercell containing the grain boundary with and without a solute atom, respectively.
$E^{\rm sp}_{\rm X}$, $E_{\rm bulk}$ are the energies of the supercell with a solute in its most stable position (hydrogen in tetrahedral site, carbon in octahedral site and boron in substitutional site) and without solute, respectively.

Then, the interaction energy between an additional solute and a boron-doped GB ($E^{\rm inter}_{\rm X-B-GB}$) has been calculated using:
\begin{equation}
    E^{\rm inter}_{\rm X-B-GB}=(E^{\rm sp}_{\rm X-B-GB}-E^{\rm sp}_{\rm B-GB})-(E^{\rm sp}_{\rm X}-E_{\rm bulk}),
\label{eqEinterDFTB}
\end{equation}
with $E^{\rm sp}_{\rm X-B-GB}$ being the energy of the supercell containing the grain boundary with boron inserted at site $I_1$ and a second solute (either boron or hydrogen), respectively.

\subsection{Results}
\label{S42}

\begin{figure}[bth!]
\centering
\includegraphics[width=0.99\linewidth]{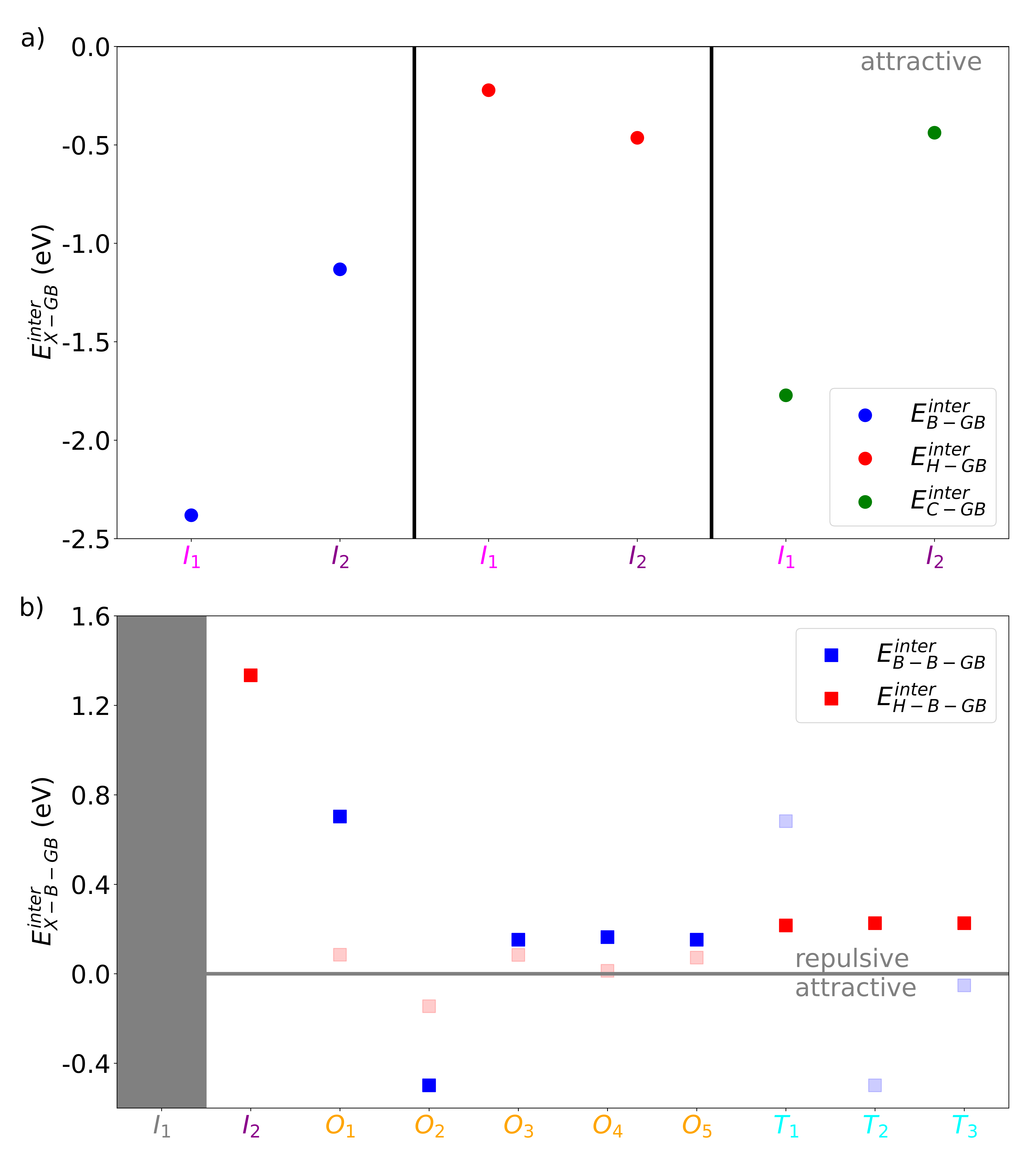}
\caption{Interaction energies between H, B, and C at boron-free and boron-doped grain boundaries. 
a) $E^{\rm inter}_{\rm X-GB}$ for boron-free GB determined using eq. (\ref{eqEinterDFT}).
b) $E^{\rm inter}_{\rm X-B-GB}$ for boron-doped GB determined using eq. (\ref{eqEinterDFTB}).
The shadowed points are situations where the solutes move to their closest stable interstitial site (tetrahedral site for hydrogen and octahedral site for boron).}
\label{FigEinterDFT}
\end{figure}

Fig. \ref{FigEinterDFT}.a presents the interaction energies of hydrogen and boron at the different sites given in fig. \ref{FigAbinitioCalcSites}.
The most stable position for boron (and carbon) in GB is in the site $I_1$ with an interaction energy of -2.38\,eV (and -1.77\,eV), but a strong attraction is also observed when both solutes are in site $I_2$. 
For hydrogen, the strongest attractive site is in the core of the GB, in $I_2$ with $E^{\rm inter}_{\rm H-GB} = -0.46$\,eV. 
This interaction energy suggests that hydrogen can be trapped in the defect with an energy up to -0.46\,eV when the GBs are similar to a $\Sigma 5(210)[001]$ GB structure. 
Since boron has the highest attraction with the grain boundaries (with an energy lower than -2\,eV), it will stay at this interface, even if other solutes like hydrogen are in its vicinity. 
These results are in agreement with the literature, where previous authors have observed the strongest attraction for boron in $\Sigma 5(310)$ GB, compared to hydrogen and carbon \cite{Kulkov2018}.

The interaction energy between solutes and boron-doped GB ($E^{\rm inter}_{\rm X-B-GB}$) is presented in fig. \ref{FigEinterDFT}.b.
In this configuration, additional boron is attracted to the GB with an interaction energy lower than -0.4\,eV in $O_2$.
This interaction energy suggests that boron can be easily accumulated in GBs, which facilitates the formation of borides when the nucleation barrier is low enough \cite{Prithiv2023,DaRosa2023}.
In contrast, hydrogen is repelled from the GB in this configuration when inserted in interstitial sites.
This result agrees with previous calculations, arguing that boron improves grain boundary cohesion and hydrogen exhibits an ion-like character in metal, including at grain boundary in $\alpha$-iron \cite{Wu1994,Kulkov2018}.
For these calculations, when hydrogen atoms are initially inserted in octahedral sites, they are moving to their closest stable tetrahedral site. 
When a second boron atom is initially inserted in the tetrahedral site, it moves towards its closest stable octahedral site.

The repulsive interaction between hydrogen and boron confirms the competition between hydrogen and boron segregation observed using TDS experiments.
Discrepancies are noted between the trapping energy from TDS measurements and the \abinitio{} calculations because the latter determines the interaction energy of one specific grain boundary type. 
In contrast, the TDS measurements determine the average trapping energy of hydrogen in all types of PAGBs.
Additionally, the calculations do not consider carbon, which also segregates to PAGBs. 
However, carbon does not saturate PAGBs when conventional heat treatments (direct quenching from austenitization temperature to ambient temperature) are performed \cite{Okada2023}.
Finally, the short-range interactions among adjacent solute atoms located in GBs are not considered for these calculations, which is an effect that could also influence the interaction energy between a solute and a defect \cite{Borges2022,Luthi2019,Hachet2020c}.
 
\section{Hydrogen concentration in steels at thermodynamic equilibrium}
\label{S5}

\begin{figure}[ht!]
\centering
\includegraphics[width=0.9\linewidth]{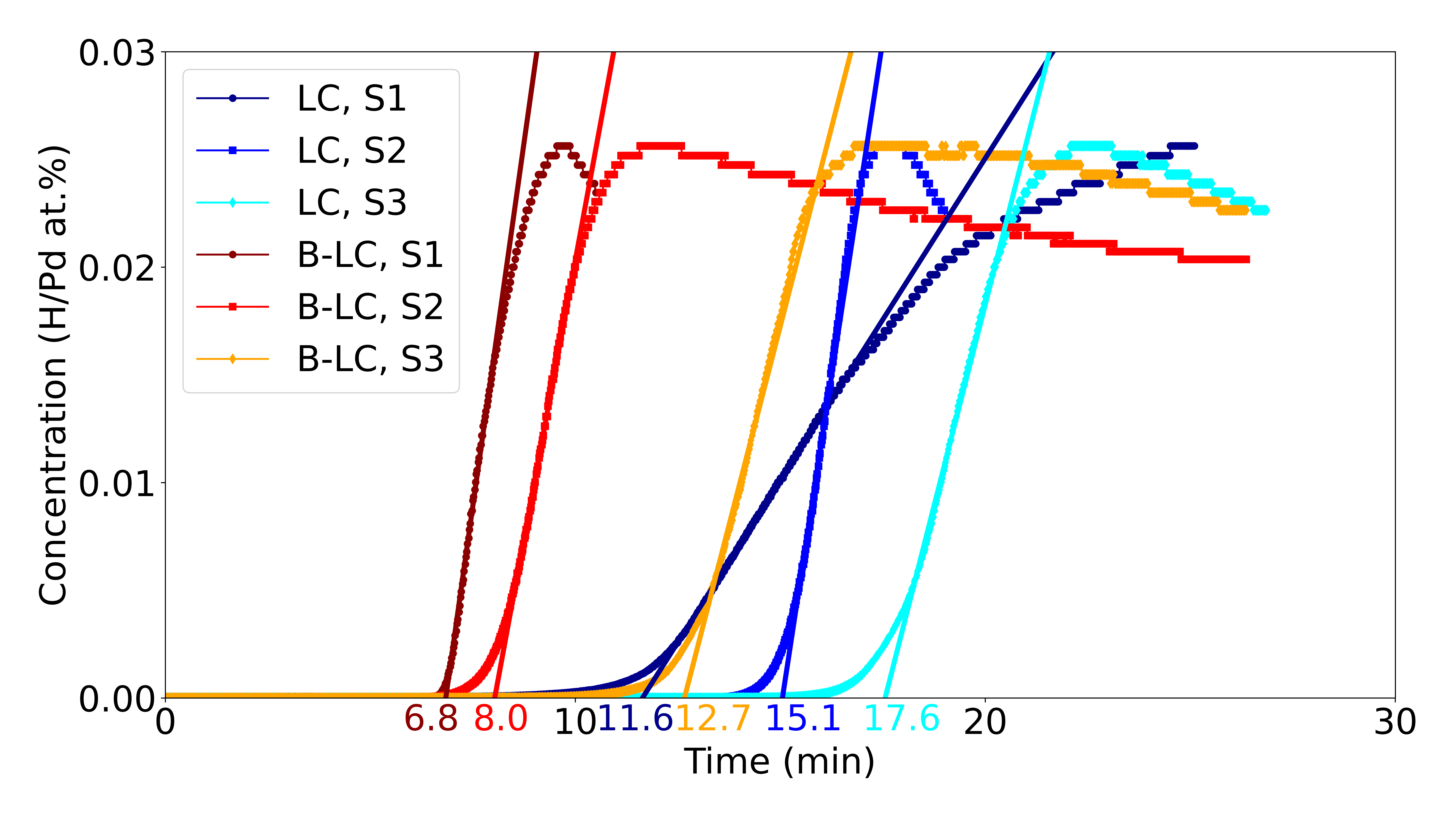}
\caption{Permeation curves of different samples (named S1, S2, and S3) for LC and B-LC steels with the lagging time ($t_{lag}$) deduced from the slope of the different curves.}
\label{FigPermTests}
\end{figure}

The previous TDS measurements show that the boron-free steel has generally an additional peak compared to the boron-doped steel and the \abinitio{} calculations show that hydrogen and boron have a repelling effect in GBs.
Consequently, it is possible to differentiate hydrogen trapped in PAGBs from hydrogen trapped in other traps using TDS measurements.
Further, the hydrogen fraction in different defects $X^{\rm H}_{\rm Defect}$ is calculated from the trapping energies determined using Kissinger's theory.
At thermodynamic equilibrium, the hydrogen segregation in crystalline defects can be approximated using the Langmuir-McLean formalism \cite{McLean1957}:
\begin{equation}
    \frac{X^{\rm H}_{\rm Defect}}{1-X^{\rm H}_{\rm Defect}}=\frac{X^{\rm H}_{\rm Bulk}}{1-X^H_{\rm Bulk}}\exp{\left(-\frac{E^{\rm tra}_{\rm Defect}}{k_{\rm B}T}\right)},
\label{eqLMIni}
\end{equation}

with $X^{\rm H}_{\rm Defect}$ and $X^{\rm H}_{\rm Bulk}$, fractions of hydrogen in a crystalline defect and the matrix, respectively.
$X^{\rm H}_{\rm Bulk}$ is calculated \via{} mass conservation from the total hydrogen concentration $C^{\rm H}_{\rm Total}$ \cite{Hachet2020c,Borges2022,Matson2021}:
\begin{equation}
    C^{\rm H}_{\rm Total} = N_{\rm L} X^{\rm H}_{\rm Bulk} + N_{\rm P_2} X^{\rm H}_{\rm P_2} + N_{\rm P_3} X^{\rm H}_{\rm P_3} + N_{\rm P_4} X^{\rm H}_{\rm P_4}.
\label{eqMatConv}
\end{equation}

with $N_{\rm L}$, $N_{\rm P_2}$, $N_{\rm P_3}$ and $N_{\rm P_4}$ being the number of possible trap sites in the lattice and from peaks 2, 3, and 4, respectively. 
The number of trap sites in the matrix is $N_{\rm L} = 6$ \cite{Krom2000,Borges2022} and $N_{\rm Defect}$ is estimated using \cite{Krom2000}:
\begin{equation}
    N_{\rm Defect} = N_{\rm L}\times \left(\frac{D_{\rm L}}{D_{\rm eff}}-1 \right) \exp{\left(\frac{E^{\rm tra}_{\rm Defect}}{k_{\rm B}T}\right)},
\label{eqNDef}
\end{equation}

where $D_{\rm L}$ is the lattice diffusion of hydrogen in $\alpha$-Fe ($D_{\rm L} = 1.45 \times 10^{-9}$ m$^{2}$.s$^{-1}$ at 298\,K \cite{Frappart2010}). 
Eq. (\ref{eqNDef}) has been chosen in this work because it is not directly dependent on the trapped hydrogen concentration and its validity was verified in previous work by Monte-Carlo simulations \cite{Kirchheim1988,Krom2000}.
The effective diffusion of hydrogen $D_{\rm eff}$ in LC and B-LC steels is determined through permeation tests at 298\,K, which are presented in fig. \ref{FigPermTests}.

The hydrogen permeation study has been conducted with a Kelvin probe-based detection method developed in-house at room temperature \cite{Evers2012,Wu2019}. 
On the side of the sample designated for hydrogen detection, a dry nitrogen atmosphere is maintained, and a 100\,nm Pd coating was applied using a Leybold Univex 450 physical vapor deposition apparatus.
The hydrogen-entry side of the sample has been subjected to the same galvanostatic condition as samples for TDS measurements.
Time lags of 11.6\,min, 15.1 and 17.6\,min are obtained for LC and 6.8\,min, 8.0 and 12.7\,min for B-LC.
These time lags give an effective diffusion coefficient of hydrogen of 5.81 $\pm$ 1.02 $\times$ 10$^{-10}$ m$^{2}$.s$^{-1}$ and 9.74 $\pm$ 2.37 $\times$ 10$^{-10}$ m$^{2}$.s$^{-1}$ for LC and B-LC, respectively.
These effective diffusion coefficients are one order higher than the ones reported in the literature (4.5 $\times$ 10$^{-11}$ m$^{2}$.s$^{-1}$ \cite{Parvathavarthini2001} and 7.4 $\times$ 10$^{-11}$ m$^{2}$.s$^{-1}$ \cite{Frappart2010}) because of the charging conditions in this present work (high current density and acidic solution with NH$_4$SCN).
It results in a higher activity of hydrogen and traps are faster filled compared to the corresponding observations reported in previous studies \cite{Parvathavarthini2001,Frappart2010}.
Additionally, since boron has a repulsive interaction with hydrogen, fewer traps need to be filled in B-LC, which explains the difference in the effective diffusion coefficient of hydrogen between LC and B-LC.
Then, the number of possible traps can be determined using equation (\ref{eqNDef}), and the results are presented in table \ref{TabNTSe}.

\begin{table}[ht]
\caption{\label{TabNTSe}Number of trap sites (in at$^{-1}$) in different microstructural defects deduced from eq. (\ref{eqNDef}) for LC and B-LC.}
\centering
\begin{tabular}{ccc}
\ & LC & B-LC \\
\hline
$N_{\rm P_2}$ & 2.52 $\times$ 10$^{-3}$ & 8.23 $\times$ 10$^{-4}$ \\
$N_{\rm P_3}$ & 1.65 $\times$ 10$^{-4}$ & 5.39 $\times$ 10$^{-5}$ \\
$N_{\rm P_4}$ & 2.35 $\times$ 10$^{-5}$ & \ \\
\end{tabular}
\end{table}

\begin{figure*}[ht!]
\centering
\includegraphics[width=0.99\linewidth]{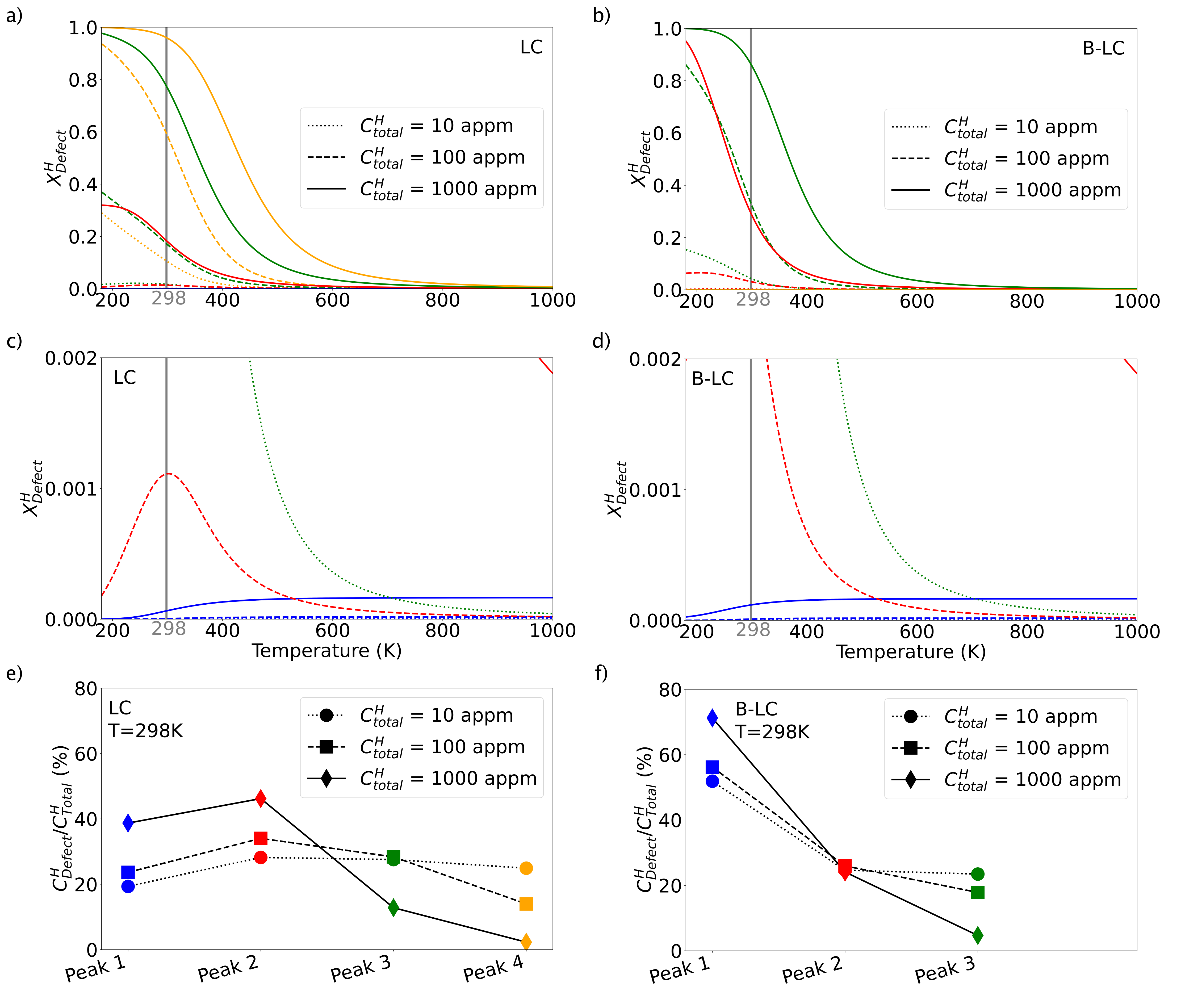}
\caption{Thermodynamic equilibrium hydrogen fraction in the different crystalline defects as a function of the temperature for total hydrogen concentration of 10\,appm (dotted lines), 100\,appm (dashed lines), and 1000\,appm (solid lines) in a) LC and b) B-LC.
The blue lines correspond to the fraction of hydrogen from peak 1 (mostly in the lattice), the red lines correspond to the fraction of hydrogen into the trapping sites of peak 2, the green lines correspond to the fraction of hydrogen into trapping sites of peak 3 and the orange lines correspond to the fraction of hydrogen in trapping sites of peak 4.
Figures c and d are zooms of figures a and b for low hydrogen fraction, respectively.
Hydrogen distribution at different trapping sites for different total hydrogen concentrations for e) LC and f) B-LC.}
\label{FigLM}
\end{figure*}

Incorporating eq. (\ref{eqMatConv}) in eq. (\ref{eqLMIni}), the latter equation can be solved self-consistently.
Fig. \ref{FigLM} presents the fraction of hydrogen in different crystalline defects for a total hydrogen concentration of 10\,appm, 100\,appm, and 1000\,appm as a function of the temperature for LC and B-LC. 

Figs. \ref{FigLM}.a-d show that all traps are not filled at thermodynamic equilibrium when the total hydrogen concentration is below 1000\,appm for the two steels.
For the LC steel, hydrogen segregates into trap sites of peak 4 because the trapping energy is lower compared to the other trapping sites.
Trap sites of peak 4 are mostly located where boron is segregating (\ie{}: PAGBs).
This observation also explains why this type of GB is the weakest microstructure feature in hydrogen environments for steels \cite{Wang2007, Okada2023}.
In addition, even if the permeation experiments revealed fewer possible trap sites ($N_{P_2}$ and $N_{P_3}$) for B-LC compared to LC, the fraction of hydrogen in trap sites 2 and 3 are higher in B-LC than in LC. 
Consequently, similar concentrations of hydrogen in traps from peaks 2 and 3 are obtained in the two steels at thermodynamic equilibrium

Then, the hydrogen distribution into the different trap sites and the lattice has been estimated in LC and B-LC and presented in figs \ref{FigLM}.e and \ref{FigLM}.f.
For LC steel, when the total hydrogen concentration is low ($C^{\rm H}_{\rm Total}$ = 10\,appm), $X^{\rm H}_{\rm Def}$ is similar for trapping sites and lattice, showing no preferential segregation.
When the total hydrogen concentration increases, traps of lowest energy are firstly saturated (\ie{}: peak 4), and the hydrogen concentration increases in other defects to possibly saturate them.
In the situation where boron protects the trapping sites of the lowest segregation energy, hydrogen is localized in the lattice with a similar concentration of hydrogen in other microstructure features. 
This effect can lead to other failure mechanisms in the materials that have been suggested in terms of specific models in the literature describing hydrogen embrittlement in metals and alloys \cite{Hirth1980,Robertson2015,Lynch2019}.

\section{Conclusion}
\label{S6}

In this work, the effect of boron segregation on the trapping behavior of hydrogen has been investigated in steel both experimentally and theoretically. 
First, the microstructure of boron-doped (B-LC) and boron-free (LC) steels has been studied using EBSD, synchrotron X-ray measurements, and correlative TKD-APT measurements to observe any change due to the presence of boron.
While the dislocation density, grain size, and grain boundary distributions are similar for LC and B-LC, pronounced segregation of boron in PAGBs is observed for the B-LC steel. 

Then, both steels have been pre-charged with hydrogen and measured through TDS measurements. 
The disappearance of one peak has been observed for B-LC compared to LC, suggesting it corresponds to hydrogen trapped in PAGBs and can accordingly be distinguished from hydrogen trapped in the other microstructural defects.

\Abinitio{} calculations demonstrate a strong attraction between boron and GBs (interaction energy of -2.38\,eV for a $\Sigma 5(210)$ GB structure), stronger than the interaction between hydrogen and GBs (interaction energy of -0.46\,eV for the same GB structure). 
Additionally, when the GB contains boron atoms, the interaction between hydrogen and this defect becomes repulsive.

Finally, the equilibrium hydrogen concentration in different defects has been approximated using the McLean formalism with the trapping energy deduced from the TDS measurements and the effective hydrogen diffusion coefficients obtained from the permeation tests. 
The latter experiment has shown that hydrogen has a higher diffusion coefficient in B-LC compared to LC because fewer traps can be filled by hydrogen in B-LC.
Different hydrogen partitioning is observed depending on temperature and total hydrogen concentration.
Compared to LC, the concentration of hydrogen in the lattice will be increased for B-LC at thermodynamic equilibrium because boron saturates the fraction of hydrogen in trap sites pertaining to PAGBs. 
This effect can induce different failure mechanisms described in different models proposed to explain hydrogen embrittlement. 
In situations where strong traps like PAGBs for hydrogen are passivated by boron, the material's resistance against hydrogen embrittlement can be modified: this hypothesis is a planned future investigation.

\textbf{Acknowledgments} -
The authors thank Dr. A. Oudriss, and Dr. E. Clouet for fruitful discussions and the help of M. Adamek, A. Sturm, and U. Tezins for technical support.
Synchrotron X-ray diffraction experiments have been conducted on Beamline P02.1, PETRA III at DESY (proposal No. I-20230183), with technical support from Dr. Alba San Jose Mendez and Dr. Alexander Schökel.
G.H. and S.L.W. acknowledge the financial support from the Alexander von Humboldt Foundation through grant n°3.3-FRA-1227460-HFST-E and n°3.1-USA-1237011-HFST-P, respectively.

\appendix
\section{EBSD and ECCI mapping of LC and B-LC}
\label{AppMicrostructure}

\begin{figure*}[ht!]
\centering
\includegraphics[width=0.8\linewidth]{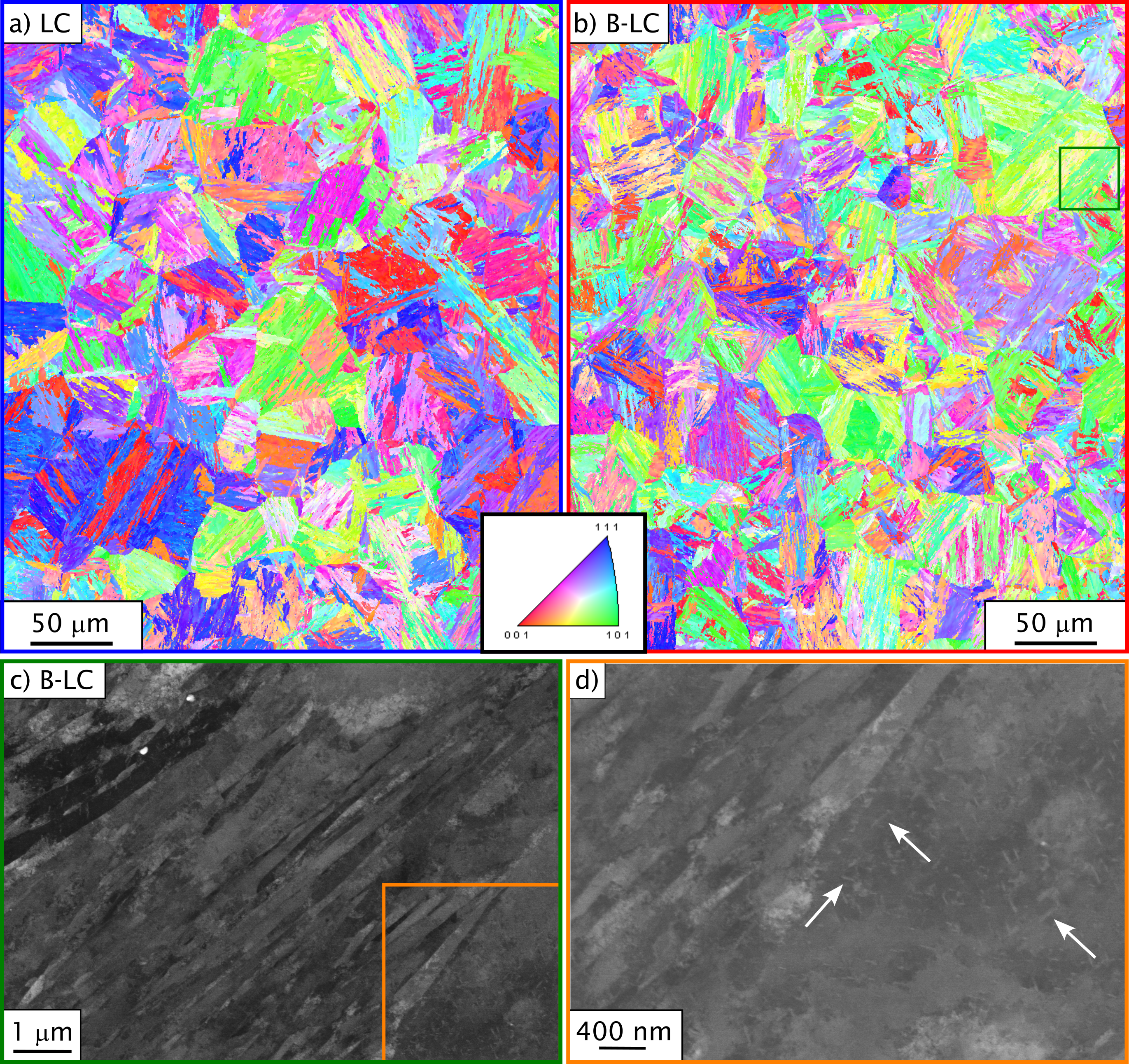}
\caption{EBSD inverse pole figure (IPF) maps of steels a) LC and b) B-LC.
c) ECCI of the area highlighted by a green square in b), showing in detail the lath microstructure, and d) carbides with all variants highlighted by white arrows.}
\label{FigMicrostructure}
\end{figure*}

The measurements of the average grain size of LC and B-LC have been performed with the EBSD maps from \ref{FigMicrostructure}.a and \ref{FigMicrostructure}.b, respectively.
For a better representation, a larger surface has been analyzed (surface of at least 2.7 $\times 10^5$ $\mu^{2}$), and a step size of 0.4\,$\mu$m has been used for the acquisition of both maps.
For the ECCI, the scanning electron microscope (SEM) was operated with an acceleration voltage of 30\,kV and a probe current of 5\,nA. 
We inverted the signal from the backscattered electron detector to obtain an image contrast similar to bright-field conditions for transmission electron microscopy images \cite{Bonnekoh2020}.

Figs. \ref{FigMicrostructure}.c and \ref{FigMicrostructure}.d show ECC images of B-LC. 
The distance between martensitic laths is around 200\,nm.
When the distance between laths becomes too large (more than 500\,nm), the formation of carbides is observed, which has been investigated using synchrotron X-ray measurements.

\section{Single and multiple events to improve the detection of boron using APT measurements}
\label{AppAPT}

\begin{figure*}[ht]
\centering
\includegraphics[width=0.99\linewidth]{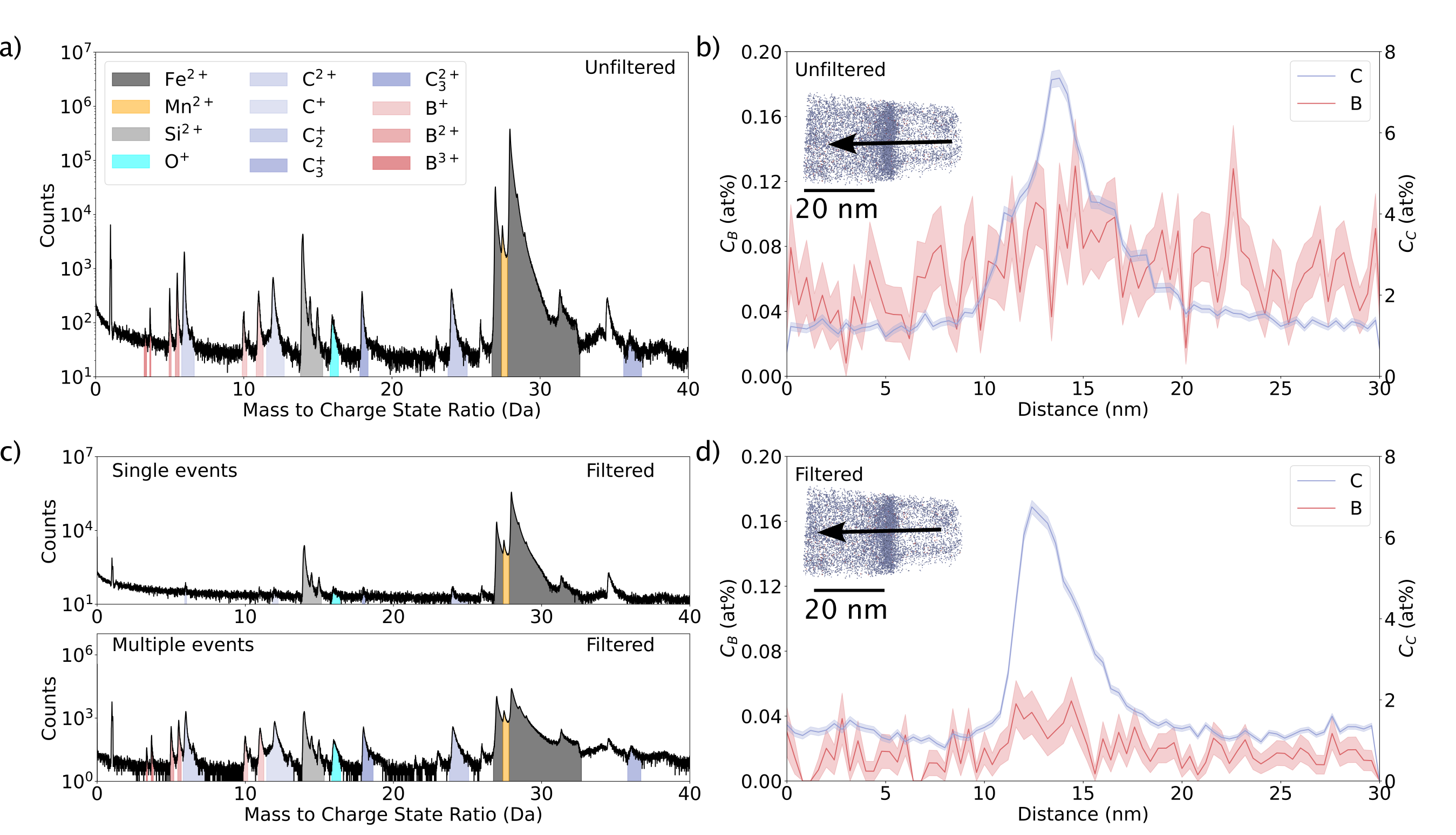}
\caption{Impact on the data filtering on the detection of boron in B-LC.
a) Mass spectrometer of the unfiltered data process with the corresponding b) 1D concentration profiles of boron and carbon
c) Mass spectra of the single event and b) multiple events detected during the same analysis with the corresponding d) concentration profile of carbon and boron.}
\label{FigAppFiltering}
\end{figure*}

The three-dimensional reconstruction presented in fig. \ref{FigAPT} has been subjected to a data filtering process based on the multiplicity of the detection events, as described by Yao \etal{} \cite{Yao2010}.
This appendix compares the concentration of carbon and boron at an interface in the boron-doped steel which has been determined with and without this filtering. 
The experiment has been carried out using a LEAP 5000 XS instrument, operated in voltage mode at 60\,K with a pulse rate of 200\,kHz, a pulse fraction of 15\,\%, and a detection rate of 50 ions per 1000 pulses.
Fig. \ref{FigAppFiltering}.a plots the mass spectrum of the specimen analyzed by APT and \ref{FigAppFiltering}.c plots the single and multiple events detected during the same experiments.
Multiple events are when a single pulse leads to multiple ions hitting the detector and have differences on the y-axis scale compared to the single events. 
Most of the carbon and boron are detected as part of multiple events and in the mass spectrum for single hits, there is barely any detectable signal for boron. 
This filter on the data improves the signal-to-background ratio specifically for boron and hence facilitates visualization of the boron distribution and quantitative analysis.
Additionally, several shoulders are observed around different peaks (\eg{}: for Fe$^{2+}$ or C$^{2+}$). 
These shoulders are mostly due to the dissociation of molecular ions, which has been discussed in previous work \cite{Peng2018}.

The corresponding one-dimensional concentration profiles of boron and carbon are then presented in figs \ref{FigAppFiltering}.b and \ref{FigAppFiltering}.d,  with and without this filtering process applied.
With this multiplicity filter used, a maximum boron concentration up to 0.045\,at\% is obtained from fig. \ref{FigAppFiltering}.d. 
However, there is barely any observable increase in boron due to the high level of background in fig. \ref{FigAppFiltering}.b, when the data are unfiltered.

\bibliographystyle{elsarticle-num-names}
\biboptions{sort&compress}
\bibliography{References}

\end{document}